\documentstyle[12pt,epsfig]{article}

\hoffset=- 0.5 truecm
\def\R{ {\rm R \kern -.31cm I \kern .15cm}}
\def\C{ {\rm C \kern -.15cm \vrule width.5pt \kern .12cm}}
\def\Z{ {\rm Z \kern -.27cm \angle \kern .02cm}}
\def\N{ {\rm N \kern -.26cm \vrule width.4pt \kern .10cm}}
\def\1{{\rm 1\mskip-4.5mu l} }

\def\noi{\noindent}

\def\lsim{\raise0.3ex\hbox{$<$\kern-0.75em\raise-1.1ex\hbox{$\sim$}}}
\def\gsim{\raise0.3ex\hbox{$>$\kern-0.75em\raise-1.1ex\hbox{$\sim$}}}
 \def\cite#1{[\ref{#1}]}

\begin{document}
\baselineskip=20 pt

\begin{center}
{\Large \bf A Sum Rule for the Two-Dimensional} \par \vskip 3
truemm
{\Large \bf  Two-Component Plasma} \par \vskip 1 truecm 
 {\bf B. Jancovici}$^1$
\end{center}
 \vskip 2 truecm  
\begin{abstract}
In a two-dimensional two-component plasma, the second moment of the {\it density}
correlation function has the simple value $\{12 \pi [1 - (\Gamma / 4)]^2 \}^{-1}$,
where $\Gamma$ is the dimensionless coupling constant. This result is derived by using
analogies with critical systems. 
\end{abstract}

\vspace{1 truecm} 

 {\bf KEY WORDS:} Coulomb systems ; sum rule ; critical behavior. \par
\vskip 3 truecm

\noindent LPT Orsay 99-59 \par
\noindent July 1999 \par
\vfill
\hbox to 4 truecm{\hrulefill}
\noi This paper is dedicated to George Stell. \par \vskip 3 truemm
\noi $^1$ Laboratoire de Physique
Th\'eorique, Universit\'e de Paris-Sud, B\^atiment 210,\par \noi 91405 Orsay Cedex,
France (Unit\'e Mixte de Recherche n$^{\circ}$ 8627 - CNRS). \par
\noi E-mail : Bernard.Jancovici@th.u-psud.fr 

\newpage
\pagestyle{plain}
\noi {\bf 1. INTRODUCTION AND SUMMARY} \\

The system under consideration is the two-dimensional two-component plasma (Coulomb
gas), i.e. a system of positive and negative point-particles of charge $\pm q$, in a
plane. Two particles at a distance $r$ from each other interact through the
two-dimensional Coulomb interaction $\mp q^2 \ln (r/L)$, where $L$ is some irrelevant
length. Classical equilibrium statistical mechanics is used. The dimensionless coupling
cons\-tant is $\Gamma = q^2/kT$, where $k$ is Boltzmann's constant and $T$ the
temperature. The system of point charges is known to be stable against collapse when
$\Gamma < 2$ and then to have the simple exact equation of state$^{(1)}$

$$p = kTn \left ( 1 - {\Gamma \over 4} \right ) \eqno(1.1)$$

\noi where $p$ is the pressure and $n$ the total number density of the particles.
\par

Let $\langle\widehat{n}(0)\widehat{n}({\bf r})\rangle_T$ be the truncated density-density
two-point function (correlation function), where $\widehat{n}({\bf r})$ is the total
microscopic number density at ${\bf r}$. The sum rule which is claimed here is

$$\int \langle \widehat{n}(0)\widehat{n}({\bf r})\rangle_T \ r^2 d^2 {\bf r} = {1 \over
12 \pi \left ( 1 -{\Gamma \over 4}\right )^2} \eqno(1.2)$$

\noi This sum rule is about the second moment of the {\it number} density correlation
function and should not be confused with the well-known Stillinger-Lovett$^{(2)}$ sum
rule obeyed by the charge correlation function

$$\int \langle \widehat{\rho}(0) \widehat{\rho}({\bf r})\rangle \ r^2 d^2{\bf r} = - {2kT
\over \pi} \eqno(1.3)$$

\noi where $\widehat{\rho}({\bf r})$ is the microscopic {\it charge} density. \par

Unfortunately, I have not been able to derive (1.2) by an argument direct enough for
my taste (perhaps the publication of the present paper will trigger somebody producing
such an argument). In Section 2, a rather indirect argument is given. In Section 3,
the sum rule is tested in two special cases. In Section 4, another indirect derivation
is made. \\

\noi {\bf 2. FROM SPHERE TO PLANE} \\

In the present Section, the sum rule (1.2) for the system living in a plane is derived
from known properties of that system living on the surface of a sphere. \\

\noi {\bf 2.1. The System on a Sphere} \\

On the surface of a sphere of radius $R$, the two-dimensional Coulomb
interaction$^{(3)}$ between two particles $i$ and $j$ at an angular distance $\psi_{ij}$
from each other is\break \noindent $\mp q^2 \ln [(2R/L)\sin (\psi_{ij}/2)]$.
Indeed, the corresponding total electric potential is a solution of Poisson's equation,
provided the total charge is zero, a condition which will be assumed. \par

The Boltzmann factor for two particles is $[(2R/L) \sin (\psi_{ij}/2)|^{\mp \Gamma}$
and the grand partition function, restricted to zero total charge, is

$$\Xi = 1 + \lambda^2R^4 \int {d\Omega_1 \ d \Omega_2 \over \left ( {2R \over L} \sin
{\psi_{12} \over 2} \right )^{\Gamma}} + \cdots \eqno(2.1)$$

\noi where $\lambda$ is the fugacity and $d\Omega_i$ is an element of solid angle
around the position of particle $i$. The last explicitly written term of (2.1)
involves 1 positive and 1 negative particles. It is easily seen that, more generally,
the term of (2.1) involving $N$ positive and $N$ negative particles is $(\lambda^2
L^{\Gamma}R^{4 - \Gamma} )^N$ times a dimensionless integral depending on $\Gamma$.
Therefore, $\ln \Xi$ depends on $\lambda$ and $R$ only through the combination
$\lambda^2 R^{4 - \Gamma}$ and obeys the homogeneity relation

$$\lambda \ {\partial \over \partial \lambda}  \ \ln \Xi = {1 \over 2} \left ( 1 -
{\Gamma \over 4} \right )^{-1} \ R \ {\partial \over \partial R} \ \ln \Xi  
\eqno(2.2)$$

\noi where the left-hand side is the total number of particles, i.e. $4 \pi R^2$ times
the total number density $n_S$ on the sphere~: 

$$4 \pi R^2 n_S = \lambda \ {\partial \over \partial \lambda } \ln \Xi \eqno(2.3)$$

For a given fugacity $\lambda$, in the large-$R$ limit, one should recover a plane
system with pressure $p$. $\ln \Xi$ must be extensive, behaving like $(p/kT)4 \pi
R^2$, and from (2.2) and (2.3) one obtains the equation of state (1.1). \par

The key ingredient of the present argument is that, for a large but not yet infinite
value of $R$, there is a universal finite-size correction$^{(4)}$ to $\ln \Xi$,
similar to the one which occurs in a system with short-range forces at its critical
point~: At a fixed fugacity, the large-$R$ expansion of $\ln \Xi$ starts as 

$$\ln \Xi = {p \over kT} \ 4 \pi R^2 - {1 \over 3} \ \ln R + \ {\rm constant} +
\cdots \eqno(2.4)$$

\noi Using (2.2) and (2.3) in (2.4), one obtains the finite-size correction to the
number density, for a given fugacity~:

$$n_S = n - {1 \over 24 \pi R^2 \left ( 1 - {\Gamma \over 4} \right )} + \cdots
\eqno(2.5)$$

\noi where $n_S$ is the number density for the system on a large sphere of radius $R$
and $n$ the number density for the plane system. \\

\noi {\bf 2.2. Stereographic Projection} \\

By a suitable stereographic projection, the two-component plasma on a sphere can be
mapped onto a modified two-component plasma on a plane. \par

Let $P$ be the stereographic projection of a point $M$ of the sphere of radius $R$ onto
the plane tangent to its South pole, from its North pole (Fig. 1). Let ${\bf r} = (x,
y)$ be the Cartesian coordinates of $P$. An area element $R^2d\Omega$ around $M$ and
its projection $d^2{\bf r}$ on the plane are related by

$$R^2d\Omega = {d^2{\bf r} \over \left ( 1 + {r^2 \over 4R^2} \right )^2} \eqno(2.6)$$

\noi The angular distance $\psi_{12}$ between two points $M_1$ and $M_2$ on the sphere
is related to the distance $|{\bf r}_1 - {\bf r}_2|$ between their projections on the
plane by

$$2R \sin {\psi_{12} \over 2} = {| {\bf r}_1 - {\bf r}_2| \over \left ( 1 +
{r_1^2 \over 4R^2} \right )^{1/2} \left ( 1 + {r_2^2 \over 4R^2} \right
)^{1/2}} \eqno(2.7)$$

Written in terms of the plane coordinates ${\bf r}_i$, the grand partition function
(2.1) on the sphere becomes

$$\Xi = 1 + \lambda^2 \int \left ( {L \over |{\bf r}_1 - {\bf r}_2|} \right )^{\Gamma}
\  {d^2{\bf r}_1 \over \left ( 1 + {r_1^2 \over 4R^2} \right )^{2 - {\Gamma \over 2}}} 
\ {d^2 {\bf r}_2 \over \left ( 1 + {r_2^2 \over 4R^2} \right )^{2 - {\Gamma \over 2}}}
\ + \cdots \eqno(2.8)$$

\noi By the same change of variables in the general term of (2.1), it can be seen that
(2.8) is the grand partition function of a modified plane two-component plasma~: in
addition to the two-body interactions $\mp q^2\ln (|{\bf r}_i - {\bf r}_j|/L)$,
there is a sign-independent one-body potential $V(r_i)$ acting on each particle~:

$${1 \over kT} \ V(r) = 2 \left ( 1 - {\Gamma \over 4} \right ) \ln \left ( 1 +
{r^2 \over 4R^2} \right ) \eqno(2.9)$$
\vskip 5 truemm

\noi {\bf 2.3. Density Shift and Linear Response} \\

A density shift is caused by the additional potential (2.9). In the large-$R$ limit,
(2.9) can be replaced by

$${1 \over kT} \ V(r) = \left ( 1 - {\Gamma \over 4} \right ) {r^2 \over 2R^2}
\eqno(2.10)$$

\noi and linear response theory gives for the density shift of the plane system at
the origin 0

$$\delta n (0) = - {1 \over kT} \int \langle \widehat{n}(0) \widehat{n}({\bf r})
\rangle_T\ V(r) \ d^2{\bf r} \eqno(2.11)$$

\noi where the statistical average in the right-hand side is to be taken in the
unperturbed system, i.e. the plane system considered in Section 1. \par

On the other hand, since the density $n_S$ on the sphere and its projection $n({\bf
r})$ on the plane are equal at the South pole 0,

$$\delta n (0) = n_S - n \eqno(2.12)$$

\noi as given by (2.5). \par

From (2.5), (2.10), (2.11), (2.12), one obtains (1.2). \\

\noi {\bf 3. TESTS} \\

The sum rule (1.2) can be tested when $\Gamma = 2$ or when $\Gamma$ is small. \\

\noi {\bf 3.1. The Case $\Gamma$ = 2} \\

At $\Gamma = 2$, the two-dimensional two-component plasma is exactly
solvable$^{(5-7)}$ by a mapping onto a free fermion field. The density correlation
function is

$$\langle \widehat{n}(0) \widehat{n}({\bf r}) \rangle_T = 2 \left ( {m^2 \over 2 \pi}
\right )^2 \left [ - K_0^2 (mr) + K_1^2(mr) \right ] +n\delta ({\bf r}) \eqno(3.1)$$

\noi where $m$ is the rescaled fugacity $m = 2 \pi \lambda L$ and $K_0$ and $K_1$
are modified Bessel functions. (3.1) does obey the sum rule (1.2), with $\Gamma =
2$. \\

\noi {\bf 3.2. The Case of Small $\Gamma$} \\

When $\Gamma$ is small, the Debye approach gives$^{(8)}$ a screened effective
potential $K_0(\kappa r)$ where $\kappa$ is the inverse Debye length~: $\kappa^2 =
2 \pi nq^2/kT = 2 \pi n \Gamma$. \par

An approximate form of the density correlation function is 

$$\langle \widehat{n}(0) \widehat{n}({\bf r}) \rangle_T = {1 \over 2} \ n^2 \left (
e^{\Gamma K_0(\kappa r)} + e^{-\Gamma K_0(\kappa r)} - 2\right )
+n\delta ({\bf r})$$

\noi or, equivalently since $\Gamma \ll 1$, the pair correlation
function is

$$h(r) = {1 \over 2} \ \Gamma^2 K_0^2
(\kappa r) \eqno(3.2)$$

\noi This approximate correlation function passes the test of the compressibility sum
rule

$$n \int h(r) \ d^2{\bf r} =
kT \ {\partial n \over \partial p} - 1 \eqno(3.3)$$

\noi since the left-hand side of (3.3) is $\Gamma / 4$ when (3.2) is used, while the
right-hand side of (3.3) is also $\Gamma /4$ at order $\Gamma$ from (1.1). \par

When used in the left-hand side of the sum rule (1.2), (3.2) gives $1/12 \pi$, which is
the correct result when $\Gamma \to 0$. \\

\noi {\bf 4. THE MASSIVE THIRRING MODEL} \\

The sum rule (1.2) can be related to another, already known, sum rule obeyed by a
field-theoretical model~: the massive Thirring model. This model is described in terms
of a two-component field $\psi$ of Dirac fermions by the Euclidean action

$$S = - \int \left [ \overline{\psi} \left ( {/ \hskip - 2.5 truemm \partial} + m_0
\right ) \psi + g (\overline{\psi} \psi )^2 \right ] d^2 {\bf r} \eqno(4.1)$$

\noi When the coupling constant $g$ does not vanish, the bare mass $m_0$ is
renormalized into $m$. \par

The massive Thirring model should obey the sum rule$^{(9)}$

$$\int \langle {\cal E}(0) {\cal E}({\bf r}) \rangle_T \ r^2 d^2{\bf r} = {1 \over 3 \pi
m^2(2 - \Delta )^2} \eqno(4.2)$$

\noi where\footnote{Ref. 9 has $i$ factors which do not appear in the present Euclidean
formalism.} ${\cal E}({\bf r}) = \overline{\psi} \psi$ and $\Delta = [1 + (g/ \pi )
]^{-1}$. This sum rule (4.2) is an application to the Thirring model of a very general
formula of Cardy$^{(10)}$ about almost-critical two-dimensional systems. In the case of
the Thirring model, (4.2) has been explicitly checked, to first order in $g$, by
Na\'on$^{(9)}$. \par

The two-component plasma can be mapped$^{(11)}$ onto the massive Thirring model, with
the correspondences

$${\Gamma \over 2} = \Delta \qquad , \qquad \widehat{n}({\bf r}) = m \ {\cal E} ({\bf
r}) \eqno(4.3)$$

\noi (the special case $\Gamma = 2$ corresponds to $g = 0$, i.e. a free-field theory,
and this is why the two-component plasma is exactly solvable at $\Gamma = 2$). The
correspondence (4.3) makes the sum rules (1.2) and (4.2) identical. \\

\noi {\bf 5. CONCLUSION} \\

The sum rule (1.2) has been derived through analogies with the theory of critical
phenomena. A more direct derivation is still wanted. \par

As a final remark, it should be noted that the two-component plasma of point particles
is stable only for $\Gamma < 2$. However, at $\Gamma = 2$, although the density $n$
diverges (for a given finite fugacity), the correlation function $\langle \widehat{n}(0)
\widehat{n}({\bf r}) \rangle_T$ is finite (for $r \not= 0$). It is tempting to
conjecture that $\langle \widehat{n}(0) \widehat{n}({\bf r})\rangle_T$ remains finite
and that the sum rule (1.2) holds in the whole range $0 < \Gamma < 4$. This
conjecture is supported by a similar statement$^{(10)}$ about the sum rule (4.2). \\

\noi {\bf 6. ACKNOWLEDGMENTS} \\

I am indebted to M. L. Rosinberg for having brought to my attention the sum
rule$^{(9)}$ (4.2), to C. M. Na\'on for stimulating e-mail exchanges, and to F. Cornu
for having encouraged me to publish the present paper. \\

\def\labelenumi{[\arabic{enumi}]}
\noindent {\bf REFERENCES} 
\begin{enumerate}
\item\label{1r} E. H. Hauge and P. C. Hemmer, {\it Phys. Norv.} {\bf 5}:209 (1971). 
   \item\label{2r} F. H. Stillinger and R. Lovett, {\it J. Chem. Phys.} {\bf
49}:1991 (1968).     
\item\label{3r} J. M. Caillol, {\it J. Physique - Lettres} {\bf 42}:L-245 (1981). 
   \item\label{4r} B. Jancovici, G. Manificat, and C. Pisani, {\it J. Stat. Phys.}
{\bf 76}:307 (1994). 
  \item\label{5r} M. Gaudin, {\it J. Physique} {\bf 46}:1027 (1985). 
\item\label{6r} F. Cornu and B. Jancovici, {\it J. Stat. Phys.} {\bf 49}:33 (1987). 
\item\label{7r} F. Cornu and B. Jancovici, {\it J. Chem. Phys.} {\bf 90}:2444
(1989).      
\item\label{8r} C. Deutsch and M. Lavaud, {\it Phys. Rev. A} {\bf 9}:2598 (1974).  
 \item\label{9r} C. M. Na\'on, {\it J. Phys. A} {\bf 22}:2877 (1989).  
  \item\label{10r} J. L. Cardy, {\it Phys. Rev. Lett.} {\bf 60}:2709 (1988). 
 \item\label{11r} See e.g. J. Zinn-Justin, {\it Quantum Field Theory and Critical
Phenomena} (Clarendon, Oxford, 1989).
  \end{enumerate}

\begin{figure}[!h] \begin{center}
    \epsfig{width=13.5cm, file=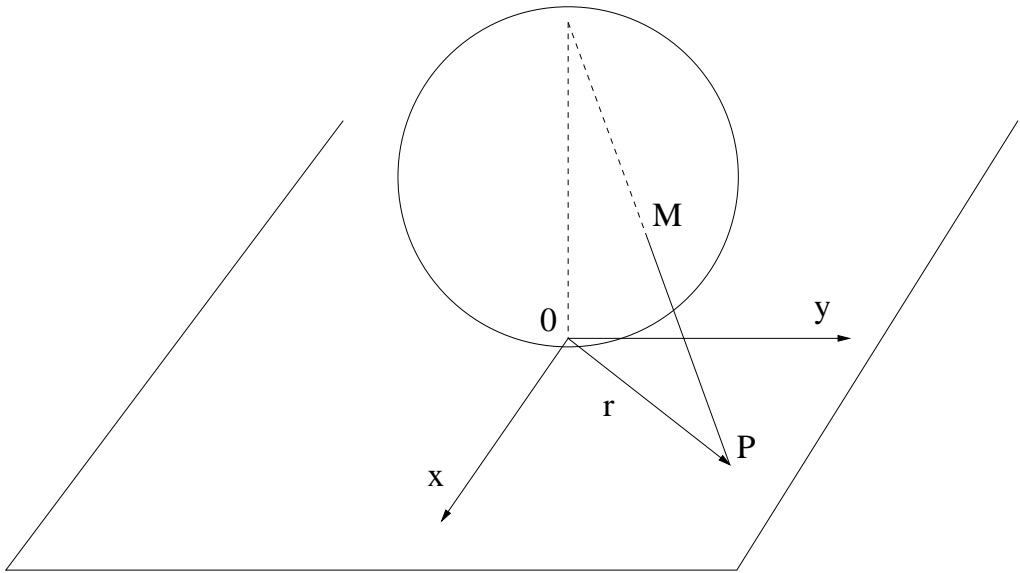}
    \caption{The stereographic projection}\end{center}
\end{figure}

\end{document}